\documentclass[12pt]{article}
\usepackage{amsfonts,latexsym,fullpage,amsmath,latexsym,amssymb}

\newtheorem{Definition}{Definition}
\newtheorem{Theorem}{Theorem}
\newtheorem{Lemma}{Lemma}
\newtheorem{Corollary}{Corollary}
\newtheorem{Remark}{Remark}
\newtheorem{example}[Theorem]{Example}

\newcommand{\onemat}[0]{{\mathbf 1}}

\newcommand{\cH}[0]{{\mathcal H}}
\newcommand{\cS}[0]{{\mathcal S}}
\newcommand{\C}[0]{{\mathbb{C}}}
\newcommand{\R}[0]{{\mathbb{R}}}
\newcommand{\N}[0]{{\mathbb{N}}}

\newcommand{\qed}[0]{\hfill $\Box$}
\newcommand{\cE}[0]{{\mathcal E}}
\newcommand{\cA}[0]{{\mathcal A}}

\title{\Large \textbf{Simulating Hamiltonians in Quantum 
Networks:\\ Efficient Schemes and Complexity Bounds}}
\author{
Pawe{\l} Wocjan\thanks{e-mail: 
{\protect\tt $\{$wocjan,roettele,janzing,eiss\_office$\}$@ira.uka.de}}, 
Martin R\"otteler, Dominik Janzing, Thomas Beth \\
\small Institut f{\"u}r Algorithmen und Kognitive Systeme,
Universit{\"a}t Karlsruhe,\\[-.5ex] 
\small  Am Fasanengarten 5, D-76\,128 Karlsruhe, Germany\\[-.5ex]
\small Forschungsgruppe Quantum Computing}

\date{September 18, 2001}

\begin{document}

\maketitle

\abstract{We address the problem of simulating pair-interaction
Hamiltonians in $n$ node quantum networks where the subsystems have
arbitrary, possibly different, dimensions. We show that any
pair-interaction can be used to simulate any other by applying
sequences of appropriate local control sequences. Efficient schemes
for decoupling and time reversal can be constructed from {\it
orthogonal arrays}. Conditions on time optimal simulation are
formulated in terms of {\it spectral majorization} of matrices
characterizing the coupling parameters. Moreover, we consider a
specific system of $n$ harmonic oscillators with bilinear
interaction. In this case, decoupling can efficiently be achieved
using the combinatorial concept of {\it difference schemes}.  For this
type of interactions we present optimal schemes for inversion.}

%
%

\section{Introduction}
The conjecture that quantum computers might be able to simulate the
time evolution of quantum systems better than classical computers has
already been stated in \cite{feynmann}. Various schemes for
constructing gate sequences which simulate the unitary evolution
corresponding to a given Hamiltonian have been suggested (see e.\,g.\
\cite{Lloyd96,SOGKL01}). More recently, a quite different approach to
this problem has become popular: considering the quantum computer as a
quantum system with Hamiltonian evolution as well, a \emph{simulation}
is a sequence of control operations acting on the quantum computer in
such a way that the net effect is a time evolution analogous to the
evolution of the system we want to simulate
\cite{dodd,graph,arrow,two,vidal,Nielsen:2001}. In the setting
described in the following the simulation problem can be stated in a
control-theoretical fashion. Assume that the total quantum system is a
quantum network, i.\,e., a system consisting of $n$ subsystems. The
Hilbert space is the tensor product
\[
\cH:= \cH_1 \otimes \cH_2 \otimes \dots \otimes \cH_n\,,
\]
where\footnote{For reasons of convenience of notation we assume that
the subsystems have equal dimensions. However, note that our results
on universal simulations generalize straightforwardly to arbitrary
dimensions.} $\cH_j:=\C^d$ for $j=1,\ldots,n$ and $d\in\N$. Let
$\{\sigma_\alpha\mid\alpha=1,\ldots,d^2-1 =: m\}$ be an orthogonal
basis of the $\R$-vector space $su(d)$ of traceless Hermitian
operators on $\C^d$. We assume that both the Hamiltonian to be
simulated and the Hamiltonian $H$ of the system are sum of
pair-interactions between nodes and free evolutions on each individual
node. Hence the system Hamiltonian is given by
\begin{equation}\label{matrix}
H:= \sum_{kl;\alpha\beta} J_{kl;\alpha\beta} \sigma^k_\alpha\sigma^l_\beta
+ \sum_{k;\alpha} r_{k;\alpha} \sigma^k_\alpha\,,
\end{equation}
where $J$ is a real symmetric $mn \times mn$-matrix and $r$ is an
$mn$-dimensional real vector. Furthermore, we assume that the only
transformations which can be implemented directly by external control
interactions are local transformations of the form
\begin{equation}\label{local}
U:=V_1\otimes V_2 \otimes\dots \otimes V_n\,,
\end{equation}
where each $V_j$ for $j=1, \ldots, n$ is an element of the special
unitary group $SU(d)$.  Assuming that the implementations of $V_j$ are
fast compared to the natural evolution given by $H$ (``fast control
limit''), the time evolution according to $U^\dagger H U$ can be
simulated by alternating the natural time evolution with
implementations of $U^\dagger$ and $U$. Here we make use of the
identity
\[
U^\dagger \exp(-iHt) U = \exp(-iU^\dagger H U t)\,.
\]
Concatenating the unitary transformations $\exp (-i U_j^\dagger H U_j
\tau_j )$ we obtain an approximation of the time evolution
corresponding to the ``average Hamiltonian''
\[
\tilde{H}:=\frac{1}{\sum_j \tau_j}\sum_j \tau_j U_j^\dagger H U_j\,,
\]
if the times $\tau_j$ are assumed to be sufficiently small (for
details see \cite{ernst})\footnote{Note that an approach of this kind
is generally accepted for describing Nuclear Magnetic Resonance
experiments.}. In this sense, the set of Hamiltonians which can be
simulated with no time overhead is exactly the {\it convex span} of
the set $\cS:= \{U^\dagger H U\}$, where $U$ is as in
eq.~(\ref{local}). For a formal definition of the notion of time
overhead of a simulation see \cite{graph,two}. Intuitively, if the
only possibility to write a Hamiltonian as a positive linear
combination of elements of the form $UHU^\dagger$ is to do it in such
a way that the sum of the coefficients is greater than one, then the
sum of these coefficients is precisely the {\it time overhead} of the
simulation. A suitably \emph{rescaled} Hamiltonian is then in the
convex span of $\cS$. As noted in \cite{graph,two}, the minimal
overhead for simulating a Hamiltonian $\tilde{H}$ by the physical
Hamiltonian $H$ is the smallest positive $\tau$ such that
$\tilde{H}/\tau $ is in the convex span of $\cS$.

Note that convex problems of this kind are closely related to the
method for obtaining pseudopure states by averaging over random
unitary transformations \cite{KCL:98}.  Pseudopure states are states
that can be written as convex combination of the maximally mixed state
(with density matrix $\onemat/d$) and a pure state $|\psi\rangle
\langle \psi|$.  Writing a general state as $\rho=\onemat/d +A$, where
$A$ is the traceless part, we have that $\rho$ can be transformed into
the pseudopure state $(1-\lambda) \onemat/d + \lambda |\psi\rangle
\langle \psi |$ by averaging over unitary transformations if and only
if $A$ can be transformed (without time overhead) into the traceless
operator $-\lambda \onemat/d +\lambda |\psi \rangle \langle \psi |$.
Determining the optimal {\it signal-to-noise ratio} of the attainable
pseudopure state is hence directly related to determining minimal
time overhead of simulation schemes.

For an arbitrary interaction between two qubits, minimization of the
time overhead has been carried out in \cite{two}. In case of $n$
qubits and pair-interaction Hamiltonians $H$ and $\tilde{H}$ the
situation is more complicated: the problem of optimal simulation of
$\tilde{H}$ by $H$ cannot be reduced to the two-qubit case. The
difficulty arising here is that the required operations on qubit $k$
for simulating an interaction between qubit $k$ and $l$ might not
coincide with the required operations on $k$ for simulating another
interaction between $k$ and $m$. The fact that the control operations
for simulating interactions between non-disjoint qubit pairs must be
consistent on the common qubit seems to be a highly non-trivial
combinatorial problem. Some upper and lower bounds for the time
overhead are given in \cite{arrow,graph}.

The issue of optimal simulation has consequences for the problem to
parallelize operations of discrete quantum algorithms. The maximal
degree of parallelization is constrained since a given interaction may
allow to implement concatenations of some two-qubit gates even if they
act on {\it non-disjoint} qubit pairs.  Assume for instance that a
Hamiltonian on three qubits is given by
\[
H:=\sigma_z\otimes \sigma_z \otimes 1 + \sigma_z \otimes 1 
\otimes \sigma_z  + 1 \otimes \sigma_z \otimes \sigma_z,
\]
where $\sigma_z$ denotes the Pauli matrix. Then, for generic times
$t>0$, the unitary $\exp (-iHt)$ is a concatenation of three two-qubit
gates, which can be implemented ``simultaneously'' by waiting the time
$t$.  But there is no obvious way for such a fast implementation of
$\exp (-iHt)$ for {\it negative} $t$ provided that $H$ is the system
Hamiltonian. One can even show that there exist negative values of $t$
such that the implementation of $\exp (-iHt)$ cannot be performed in
principle within a time period of length $|t|$, since $H$ is not able
to simulate $-H$ without time overhead \cite{arrow}. For $n$ qubits
with $\sigma_z \otimes \sigma_z$-interactions the minimal overhead is
known to be at least $n-1$. This shows that the question of optimal
parallelism of sequences of gates is rather sensitive to the form of
the underlying Hamiltonian. Hence the question of optimal
implementation of complex transformations on an $n$-partite system
cannot be answered alone by resolving the network into elementary
gates: the question of maximally possible parallelism appears already
on the control-theoretic level.

The paper is organized as follows. In Section~\ref{network} we
construct decoupling schemes from orthogonal arrays.  Selective
decoupling can be achieved by straightforward generalizations.
Therefore each pair-interaction Hamiltonian on an $n$-partite systems
can be converted into any other provided that a sufficient set of
local unitary control operations is available (the so-called {\it
transformer groups} introduced in \cite{WRJB01}). Such a simulation
has time overhead of $O(n^2)$. Our schemes works even if the
dimensions of the $n$ subsystems do not agree, thereby solving a
problem stated in \cite{Nielsen:2001}.

A {\it lower} bound on the time complexity of mutual simulation of
different Hamiltonians is shown which uses the spectra of matrices
describing the coupling parameters of the interaction.  Efficient
schemes for switching off all the interactions (``decoupling
schemes'') are constructed. Using the combinatorial concept of {\it
orthogonal arrays} decoupling can be achieved if local transformations
of a unitary error basis are available. The length of the required
sequence of local transformations grows linearly with the number of
nodes.

A specific form of interactions is dealt with in
Section~\ref{harmonic}. There we consider harmonic oscillators coupled
by bilinear terms of creation and annihilation operators. In this case
rather simple decoupling schemes can be constructed on the basis of
{\it difference schemes}, another concept from combinatorics.
Furthermore we discuss the generalization of this problem where a
given bilinear interaction is used to simulate another interaction of
this form with different coupling parameters. Finding the time optimal
simulation is equivalent to the mathematical problem of constructing
vectors of minimal length with complex entries of modulus one such
that their inner products yield certain values.  This fact is used for
deriving a lower bound on the simulation time. In contrast to the
general system Hamiltonian, the {\it upper} bound is $O(n)$.  An
optimal scheme for time inversion is constructed.

In Section~\ref{compare} we discuss the connection between the
decoupling schemes of Section~\ref{network} and the schemes presented
in \cite{leung}.

%
%

\section{Simulating Hamiltonians in Networks}\label{network}

\subsection{Selective Decoupling}

A useful tool for simulating Hamiltonians in multi node quantum
networks is given by decoupling schemes, i.\,e., sequences of local
operations which switch off unwanted interactions.  First we describe
schemes for decoupling all the interactions. Schemes for decoupling
two interacting quantum systems---mainly in the context of a quantum
register which is coupled to a bath---have been derived in
\cite{violaDecoupling,Zanardi00}). Decoupling of $n$-partite systems
interacting with each other by two-body Hamiltonians has been
considered in \cite{leung, stoll} for the case of qubits. In
\cite{Nielsen:2001} it was noted that decoupling of bipartite systems
can be concatenated in such a way that the system is separated into
suitable clusters of subsystems without coupling between different
clusters. The number of operations involved in this scheme is
$O(n^{2\log_2 d})$. In this paper we present a construction based on
orthogonal arrays which uses $O(n)$ operations provided that the
dimension $d$ of the nodes is a prime power.

Straightforward generalizations of decoupling schemes allow to switch
off all interactions except the Hamiltonian of a pair of nodes
(``selective recoupling'') or the Hamiltonian of a single node.  We
have shown in \cite{WRJB01} that any bipartite Hamiltonian can
simulate any other provided that it consists of non-trivial local
Hamiltonians on both nodes and a non-trivial coupling. If this
criterion is met by all pairs of an $n$-partite Hamiltonian then
universal simulation of all pair-interaction Hamiltonians is
possible\footnote{Note that the condition that {\it all} the couplings
have to be non-trivial is only necessary in the average Hamiltonian
approach. If higher order terms in the time interval are considered
interactions between nodes $k$ and $l$ resp. $k$ and $m$ can be used for
simulating a coupling between $l$ and $m$ as noted in
\cite{dodd,Nielsen:2001}.}.  This result is also true if not every
local operation is available, it is sufficient that all the elements
of a so-called {\it transformer group} (a concept introduced in
\cite{WRJB01}) can be implemented.  The simulation time overhead is of
order $O(n^2)$ since there are $n(n-1)/2$ pairs of nodes.

Rephrasing well-known results about decoupling
(e.\,g. \cite{violaDecoupling,Zanardi00}) in our language, we briefly
describe decoupling for bipartite systems. Remember that the
Hamiltonian on a $d$-dimensional quantum system can be cancelled if an
appropriate sequence of unitary control operations is applied
\cite{WRJB01}.

\begin{Definition}[Annihilator]
An annihilator $A:=(U_1,\tau_1,U_2,\tau_2,\ldots,U_N,\tau_N)$ of
dimension $d$ and length $N$ is given by unitaries $U_i\in SU(d)$
and relative times $\tau_i>0$, $\sum_{i=1}^N \tau_i=1$ such that
\[
\sum_{i=1}^N \tau_i U_i^\dagger a U_i = 0 
\]
for all $a\in su(d)$. An annihilator is called minimal if there is no
shorter annihilator.
\end{Definition}
A minimal annihilator of dimension $d$ has length $d^2$ (as already
noted in \cite{violaDecoupling} and proved in \cite{WRJB01}) and all
$\tau_i$ are equal.  Moreover, the unitaries $U_i$ must form a {\it
unitary error basis} of operators in $\mathbb{C}^{d\times d}$, i.\,e.\
a collection of $d^2$ unitaries $U_i$ that are orthogonal with respect
to the trace inner product $\langle A|B\rangle:=\mathrm{tr}(A^\dagger
B)/d$. Such unitaries can be explicitly constructed using nice error
bases \cite{KR:2000a,KR:2000b}.

Being a special case of equation (\ref{matrix}) we write the general
Hamiltonian of a bipartite system as follows:
\begin{equation}
H=\sum_{\alpha\beta} J_{\alpha\beta}\,\sigma_{\alpha}\otimes\sigma_{\beta} +
\sum_\gamma r_\gamma\, \onemat\otimes \sigma_\gamma +
\sum_\delta s_\delta\, \sigma_\delta \otimes \onemat\,.
\end{equation}
Let $\mathcal{E}_1=\{U_i\}$ and $\mathcal{E}_2=\{V_i\}$ be unitary
error bases of the respective systems and let $\mathcal{A}$ denote the
set $\{1,\ldots,d^2\}$. By applying the annihilators defined by
$\mathcal{E}_1$ and $\mathcal{E}_2$ independently on the nodes we can
switch off the Hamiltonian, i.\,e.\
\begin{equation}\label{switchofftwo}
\frac{1}{|\mathcal{A}^2|}\sum_{(i,j)\in\mathcal{A}^2} 
(U_i^\dagger \otimes V_j^\dagger)\, H\, (U_i\otimes V_j) = 0\,.
\end{equation}

We describe a decoupling scheme on $n$ nodes by $n$ unitary error
basis $\cE_1,\ldots,\cE_n$ and an $n\times N$-matrix
$M=(m_{ij})_{i=1,\ldots,n, j=1, \ldots, N}$. This matrix contains
elements of $\mathcal{A}$ specifying the conjugation by the
$m_{ij}^{\rm{th}}$ unitary of $\cE_i$ on a specific node $i$ for a
certain time interval $j$. The $N$ time intervals correspond to the
columns and the different nodes correspond to different rows. For
instance the decoupling scheme corresponding to (\ref{switchofftwo})
is given by the array
\begin{equation}
\left(\begin{array}{cccc|cccc|c|cccc}
1 & 1 & \ldots & 1 & 2 & 2 & \ldots & 2 & \ldots & N & N & \ldots & N\\
1 & 2 & \ldots & N & 1 & 2 & \ldots & N & \ldots & 1 & 2 & \ldots & N\\
\end{array}\right)
\end{equation}
and unitary error bases $\cE_1=\{U_i\}$ and $\cE_2=\{V_i\}$
corresponding to the first and second node, respectively.

The simplest approach for decoupling is to choose the columns of $M$
as all tuples of $\mathcal{A}^n$. However, this scheme is
not efficient in terms of the number of time intervals and pulses
since both scale exponentially as $d^{2n}$ because the sequence has to
be repeated $d^2$ times for each added node.

More efficient schemes can be constructed using the combinatorial
structure of orthogonal arrays. See \cite{BJL1,CD:96,sloane} for the
general theory of orthogonal arrays. Orthogonal arrays have numerous
applications e.\,g., in the design of experiments. Also there are
connections between orthogonal arrays and mutually orthogonal Latin
squares and transversal designs (cf. \cite[Section VIII]{BJL1}). The
following definition takes account of the fact that for purposes of
decoupling we need a special type of orthogonal arrays, namely those
of strength $t=2$ (cf. \cite{BJL1,CD:96,sloane} for the general
case). Also the notation used is adapted to this situation.

\begin{Definition}
Let ${\cal A}$ be a finite alphabet and let $n, N \in \N$. An $n\times
N$ array $M$ with entries from $\mathcal{A}$ is an orthogonal array
with $|\mathcal{A}|$ levels, strength $t=2$, and index $\lambda$ if
and only if each pair of elements of $\mathcal{A}$ occurs $\lambda$
times in the list $((m_{kj},m_{lj})\mid j=1,\ldots N)$ for $1\le
k<l\le n$. We use the notation $OA_\lambda(n,N)$ to denote a
corresponding orthogonal array.\footnote{Note that in \cite{BJL1} the
notation $OA_\lambda(n, s)$ is used for an orthogonal array with
$N=\lambda s^2$ in our notation, where $s := |{\cal A}|$.}
\end{Definition}

The following theorem shows that decoupling in networks of arbitrary
dimensions can be achieved using pulse sequences obtained by
orthogonal arrays. 

\begin{Theorem}[Decoupling]\label{decoupOA}
Let ${\cal A}$ be the finite alphabet
$\{1, \ldots, d^2\}$. Then any orthogonal array with parameters
$OA_\lambda(n,N)$ over ${\cal A}$ can be used to decouple a quantum
network consisting of $n$ nodes of dimension $d$. The number of local
operations used in this scheme is given by $N$.
\end{Theorem}
{\bf Proof:} Let $M=(m_{ij})$ be an $n\times N$-matrix over
$\mathcal{A}$ corresponding to the parameters $OA(n,N)$. Choose
unitary error bases ${\cal E}_1, \ldots, {\cal E}_n$ where ${\cal E}_k
= \{E^k_i : i = 1, \ldots, d^2\}$ to define annihilators for each
node. For each row $k$ of $M$ let $(E^k_{m_{k, j}} : j = 1, \ldots,
N)$ be the corresponding local conjugations on node $k$.  We now
consider a pair of nodes, i.\,e., two rows $k$ and $l$ of $M$, and
show that the local terms and the coupling between the two nodes are
switched off. Since each pair of elements of $\mathcal{A}$ occurs
precisely $\lambda$ times in the list $((m_{kj},m_{lj})\mid
j=1,\ldots,N)$ the averaged Hamiltonian $H_{k,l}$ on the nodes $k$ and
$l$ is given by (setting $U_i := E^k_i$ and $V_i := E^l_i$)
\[
\frac{1}{|\mathcal{A}^2|}\sum_{(i,j)\in\mathcal{A}^2} 
(U_i^\dagger \otimes V_j^\dagger)\, H_{k,l}\, (U_i\otimes V_j).
\]
This sum is equal to zero since both annihilators are applied
independently on both nodes. 
\qed

For any given number $n\in \N$ of nodes there are parameters
$\lambda,N$ such that an orthogonal array $OA(n,N)$ exists. However,
since we are interested in efficient schemes, $N$ has to be a
polynomial in the number $n$ of nodes. Also it is of interest to give
explicit constructions of such schemes, i.\,.e, of orthogonal arrays.
Whereas little is known about the existence of efficient schemes for
general $n$ and alphabet size $s := |{\cal A}|$ the situation is much
better in the case when $s$ is a prime power. 

\begin{Corollary}\label{decoupNetwork}
Let an $n$-node quantum network with pair-interaction Hamiltonian be
given and let the dimension $d$ of each node be a prime power. Then
there exists a decoupling scheme using $N$ local operations, where
$N\leq c n$ and $c$ is a constant depending only on $d$. 
\end{Corollary}
{\bf Proof:} Let $s:= d^2$ be the size of a minimal annihilator for a
$d$-dimensional system. In view of Theorem \ref{decoupOA} we have to
show that there exists an orthogonal array with parameters
$OA_\lambda(\tilde{n}, N)$ with $n \leq \tilde{n}$ and $N \leq c n$ as
above. The result \cite[Theorem 3.20]{sloane} gives an explicit
construction for an $OA((s^i-1)/(s-1), s^i)$ for any $i \geq
2$. Hence, if for the number of nodes $n=(s^i-1)/(s-1)$ holds, we have
found a decoupling scheme with $N=s^i = (s-1) n + 1$ operations,
i.\,e., $N = O(n)$. For general $n$ we embed into an OA of this
form. Switching to the next number of the form $(s^i-1)/(s-1)$ with
suitable $i\geq 1$ can be achieved be multiplying $n$ with a number
less or equal $s$, i.\,e., $\tilde{n} \leq s n$.  \qed

\begin{Remark}
There are tables of OAs covering the small instances
(cf. \cite{BJL1,CD:96,sloane}). We remark that there is a family of
OAs with parameters $OA(2s^i, 2\frac{s^i-1}{s-1}-1)$ \cite[Theorem
6.28]{sloane}. This shows that the constant $c$ in Corollary
\ref{decoupNetwork} can be improved to $c/2$.
\end{Remark}

The following example illustrates that OAs give more efficient schemes 
already for small systems. 

\begin{example}
We consider the case of four three-level systems. Using the
exponential scheme we need $s^4=6561$ local operations to decouple all
interactions. Following \ref{decoupNetwork} we obtain a decoupling
scheme with the same property that uses only $s^2=81$ local
operations.
\end{example}

Selective decoupling can be achieved as follows. If the decoupling
scheme is applied to all but one or two nodes, the  remaining
Hamiltonian is the local Hamiltonian of the node or the bipartite
Hamiltonian of the two nodes, respectively.

The assumption that every node is coupled to all the other nodes is
too strong in many physical systems since many coupling terms might be
neglected.  This reduces the overhead for decoupling and inverting the
time evolution.

In general the interaction graph of a partially coupled network is a
non-complete graph. Each graph can be colored, by assigning each
vertex one of a number of different colors. Such a coloring scheme is
called a {\em proper} coloring if no two connected vertices have the
same color. The chromatic number $\chi$ is the smallest number of
colors required to properly color the graph. In a complete graph (a
fully coupled network) $\chi=n$, but in a partially coupled network
the chromatic number can be much smaller. This observation permits to
derive more efficient decoupling schemes \cite{knill} since if the
network is represented by a properly colored graph, then there are no
constraints on the pulse sequences between nodes with the same
color. It is sufficient to create a decoupling scheme of a completely
coupled $\chi$-node network, and apply identical sequences to all
nodes of the same color.

Note that the selective decoupling scheme presented above generalize
straightforwardly to the case that the dimensions of the $n$
subsystems do not agree. Then one has to use so-called {\em mixed
orthogonal arrays}, i.\,e., one has different alphabets $\cA$ for
different nodes.  Although little is known about constructions of
efficient mixed orthogonal arrays, it is known that exponential ones
exist (cf. \cite[Section 9.3]{sloane}).  Furthermore, the mutual
simulation of Hamiltonians on bipartite systems is also possible for
different dimensions, since the proof in \cite{WRJB01} does only rely
on the fact that on each of the subsystems (possibly different)
transformer groups are available.

\subsection{Lower bound on the simulation time overhead}

In order to derive lower bounds on the overhead for the simulation
time we neglect the free evolution and consider the weaker problem to
simulate the desired Hamiltonian on $n$ nodes up to local terms of
each node. Note that the local terms become irrelevant when allowing
arbitrary unitary operations on the nodes.  

In the following it will be convenient to represent the interaction
Hamiltonian by the so-called $J$-matrix
\begin{equation}\label{Jmatrix}
J=\left(
\begin{array}{ccccc}
0      & J_{12} & J_{13} & \cdots & J_{1n} \\ 
J_{21} & 0      & J_{23} & \cdots & J_{2n} \\ 
J_{31} & J_{32} & 0      &        & J_{3n} \\ 
\vdots & \vdots &        & \ddots &        \\ 
J_{n1} & J_{n2} & J_{n3} & \quad & 0
\end{array}
\right)\in\mathbb{R}^{m n\times m n}\,,
\end{equation}
where the matrix $J_{kl}$ describes the coupling between the nodes $k$
and $l$ and $J_{lk}$ is the transpose of $J_{kl}$ in eq. (\ref{matrix}).

Let $H$ and $\tilde{H}$ be arbitrary pair-interaction Hamiltonians. We
investigate the question whether $\tilde{H}$ can be simulated by $H$
with overhead $\tau$. This defines a quasi-order of the
pair-interaction Hamiltonians for $\tau=1$. A partial characterization
of the quasi-order is expressed in terms of majorization of the
spectra of the corresponding matrices $J$ and $\tilde{J}$.  Note that
this criterion does not coincide with the criterion given in
\cite{two} for two qubits since the latter refers to the spectrum of
the Hamiltonians and not of the $J$-matrices. To make this more
precise we introduce the following notation.  Let $x=(x_1,\ldots,x_d)$
and $y=(y_1,\ldots,y_d)$ be two $d$-dimensional real vectors. We
introduce the notation $\downarrow$ to denote the components of a
vector rearranged into non-increasing order, so
$x^\downarrow=(x_1^\downarrow,\ldots,x_d^\downarrow)$, where
$(x_1^\downarrow\ge x_2^\downarrow\ge\ldots\ge x_d^\downarrow)$. We
say that $x$ is majorized by $y$ and write $x\prec y$, if
$$ \sum_{j=1}^k x_j^\downarrow\le \sum_{j=1}^k y_j^\downarrow\,, $$
for $k=1,\ldots,d-1$, and $\sum_{j=1}^d x_j^\downarrow = \sum_{j=1}^d
y_j^\downarrow$ (see \cite{bha}). The usefulness of majorization
techniques in quantum information processing has first been
demonstrated in \cite{Nielsen:99} in the context of transforming
states under local operations and classical communication. In
\cite{graph} majorization was used in the context of simulation of
Hamiltonians in quantum networks of qubits.

Let $\mathrm{Spec}(X)$ denote the spectrum of the Hermitian matrix
$X$, i.\,e.\ the vector of eigenvalues. Ky Fan's maximum principle
\cite{bha} gives rise to a useful constraint on the eigenvalues of a
sum $C:=A+B$ of two Hermitian matrices: 
\begin{equation}\label{majorsum}
 \mathrm{Spec}(A+B)\prec \mathrm{Spec} (A)+ \mathrm{Spec}(B)\,.
\end{equation}
Based on these tools we derive the following bound, which generalizes
the bounds given in \cite{graph} from qubits to qudits. The second
statement of the theorem even tightens the bound for qubits given
there.

\begin{Theorem}[Lower bound]\label{Scale}
Let $H$ and $\tilde{H}$ be arbitrary pair-interaction Hamiltonians. A
necessary condition that $\tilde{H}$ can be simulated with overhead
$\tau$ by $H$ is that the spectrum of $\tilde{J}$ is majorized by the
spectrum of $\tau J$.  Furthermore it is necessary that this
majorization criterion is still satisfied after rescaling the
couplings as follows: $J_{kl}':= s_{kl} J_{kl}$ and
$\tilde{J}':=s_{kl} \tilde{J}$, where $S=(s_{kl})$ is an arbitrary
real symmetric $n\times n$-matrix.
\end{Theorem}
{\bf Proof:} We choose an orthonormal basis $B$ (with respect to the
trace inner product) for $su(d)$. Each $a\in su(d)$
can be represented by $|a\rangle=(tr(a\sigma_1),\ldots
tr(a\sigma_{d^2-1}))^T$ with $\sigma_i\in B$.  Conjugation of $a\in
su(d)$ by a unitary $u\in SU(d)$ corresponds to an element of
$SO(d^2-1)$ rotation of $|a\rangle$. The subgroup of $SO(d^2-1)$
defined by the adjoint action of $SU(d)$ in this way will be denoted
by ${\mathcal R}$.

Representing the Hamiltonians $H$ and $\tilde{H}$ by their $J$
matrices we see that $\tilde{H}$ can be simulated with overhead $\tau$
if and only if there is a sequence of special orthogonal matrices
$U_j=U_{j1}\oplus U_{j2}\oplus\cdots\oplus U_{jn}$, where $U_{j,k} \in
{\cal R}$ for all $j, k$ and $\tau_i>0$ with $\sum_j\tau_j=\tau$ such
that
\begin{equation}\label{mixing}
\tilde{J}=\sum_j \tau_j U_j J U_j^T\,.
\end{equation}
The proof now follows from Uhlmanns's theorem \cite{vidalmaj} which
implies that spectrum $\tilde{J}$ is majorized by the spectrum of
$\tau J$.

The second statement is a direct consequence from the first one since
the same simulation procedure can be used for the rescaled problem.
\qed

Note that for $d>2$ not every rotation in $SO(d^2-1)$ corresponds to a
conjugation by a unitary in $SU(d)$. Therefore this condition seems to
be the weaker the greater $d$ gets. A necessary condition for the fact
that $H$ can simulate $\tilde{H}$ with overhead $\tau$ is that
$\mathrm{Spec}(\tilde{H})\prec\tau \mathrm{Spec}(H)$, i.\,e.\ a
condition based on majorization of the spectra of the
Hamiltonians. Clearly the size of the matrices representing the
Hamiltonians grows exponentially with $n$, whereas the size of their
$J$-matrices grows only linearly with $n$. This makes the
$J$-representation useful for calculations.  Furthermore, in some
important cases it permits to derive tight bounds (see Section
\ref{harmonic} and \cite{arrow}).

\subsection{Inverting}

We now consider the problem to invert an arbitrary, possibly unknown
Hamiltonian in a quantum network, i.\,e.\ to simulate $-H$ when $H$ is
present.  It is well-known that this question is closely related to
the construction of decoupling schemes
(e.\,g. \cite{RPW:70,violaDecoupling,leung}). Lower bounds on the time
overhead for time-reversal in $n$-qubit systems were given in
\cite{arrow}. In the case of a single node we can invert the time
evolution by summing over all elements of the unitary error basis but
the identity. This trick can be generalized to the case of multiple
nodes. For that we introduce the notion of \emph{normal form} for
decoupling schemes.

\begin{Definition}[Normal form]
Let ${\cal O}$ be an $OA_\lambda(n,N)$. Assume in addition that the
alphabet ${\cal A}$ consists of the elements of a finite group $G$. We
say that ${\cal O}$ is in normal form if each entry in the first column of
${\cal O}$ is the identity element of $G$.
\end{Definition}

\begin{Lemma}\label{normalform}
Let ${\cal O}$ be an $OA_\lambda(n,N)$. Then there is an orthogonal array
with the same parameters which is in normal form.
\end{Lemma}
{\bf Proof:} We identify the underlying alphabet ${\cal A}$ with an
arbitrary finite group $G$ of order $|{\cal A}|$. Consider two rows
$(g_1, \ldots, g_N)$ and $(h_1, \ldots, h_N)$ of ${\cal O}$. Multiplying the
elements of the rows by $g_1^{-1}$ and $h_1^{-1}$ respectively
preserves the property that all pairs occur with frequency $\lambda$
since $G \times G$ is invariant under multiplication by fixed
elements.  
\qed

Based on the normal form of OAs we now give an inversion scheme for a
general, possibly unknown pair-interaction Hamiltionian.

\begin{Theorem}[Inverting]
Any $OA(n,N)$ can be used to invert the time evolution of a quantum
network consisting of $n$ nodes. The number of local operations used
is $N-1$ and the time overhead is $N-1$.
\end{Theorem}
{\bf Proof:} By Lemma \ref{normalform} we assume that the orthogonal
array ${\cal O}$ is in normal form and $M$ is the corresponding
$n\times N$ matrix over the alphabet ${\cal A}$ which in turn is
identified with the elements of a finite group $G$. Furthermore, we
choose unitary error bases ${\cal E}_1, \ldots, {\cal E}_n$ where
${\cal E}_k = \{E^k_i : i = 1, \ldots, d^2\}$ and identify the
identity element of $G$ with the corresponding identity matrices. For
any pair $k$, $l$ of nodes we obtain the following identity for the
averaged interaction $H_{k,l}$ between $k$ and $l$:
\[
-H_{k,l}= 
\sum_{j=2}^N 
(E^k_{m_{kj}} \otimes E^l_{m_{lj}})^\dagger H_{k,l}
(E^k_{m_{kj}} \otimes E^l_{m_{lj}}).
\]
Since the interactions between all nodes are inverted it follows that
$H$ is inverted. 
\qed

Similar to the results given in \cite{arrow} we derive a lower bound
for the time overhead for inverting a dynamical evolution as follows.

\begin{Lemma}[Lower bound on inverting]\label{invertLower}
Let $r$ be the greatest eigenvalue and $q$ the smallest eigenvalue of
the $J$-matrix representing $H$. Then $\mu\ge\frac{r}{-q}$ is a lower
bound on the overhead for simulating $-H$ by $H$.
\end{Lemma}
{\bf Proof:} Denote the smallest eigenvalue of a matrix $A$ by
$\lambda_{\min} (A)$. Then we have
\[
-r=\lambda_{\min}(-J)=\lambda_{\min} \Big(\sum_j \tau_j U_j J U^T_j\Big)\geq
\tau\lambda_{\min} (J)= \tau q \,.
\] 
The inequality is due to
$\lambda_{\min}(A+B)\ge\lambda_{\min}(A)+\lambda_{\min}(B)$ for the
sum of two Hermitian matrices $A$ and $B$ (see \cite{bha},
Theorem~III.2). Since $q$ is negative ($J$ is traceless) we have
$\tau\ge -r/q$. 
\qed

Consider a fully coupled quantum network, where each pair-interaction
is of the form $\sigma_\alpha \otimes \sigma_\alpha$. Then the
greatest and smallest eigenvalues of $J$ are $n-1$ and
$-1$. Lemma~\ref{invertLower} gives $n-1$ as lower bound on the time
overhead for inversion.

%
%

\section{Harmonic oscillators with bilinear couplings}\label{harmonic}

We consider a quantum network, where the $n$ nodes are harmonic
oscillators, i.\,e., their energy values are given by
$E=0,1,2,\dots$. Here we restrict our attention to the case of energy
values less than $d$ and obtain as an approximation a $d$-dimensional
Hilbert space for each oscillator.  The Hamiltonian of the uncoupled
system is given by
\begin{equation}
H_0:= \sum_{k=1}^n h_k\,,
\end{equation}
where each $h_k$ is the diagonal matrix ${\rm diag}(0,1,\dots,d-1)$ on
the node $k$. Assume in addition that there is a coupling $H_C$ between
the individual harmonic oscillators having the form
\begin{equation}\label{harmonicCoupling}
H_C:= \sum_{k,l}c_{kl} a_k  a^\dagger_l 
\end{equation}
Here $C=(c_{kl})$ is a real symmetric $n\times n$-matrix with zeros on
the diagonal determining the couplings and $a_k$ and $a_l^\dagger$ are
the \emph{annihilation} and \emph{creation} operators on the
$k^{\rm{th}}$ and $l^{\rm{th}}$ oscillator, respectively. The
annihilation operator $a$ is defined by
\begin{equation}
a|0\rangle=0\,,\quad\quad a |E\rangle = \sqrt{E} |E-1\rangle\,,
\quad E=1,2,\ldots,d-1\,,
\end{equation}
where $\{|E\rangle\mid E=0,1,\ldots,d-1\}$ is an eigenvector basis of the
free Hamiltonian of an harmonic oscillator.

Interactions of the form~(\ref{harmonicCoupling}) often appear if
higher order terms in creation and annihilation operators are
neglected and only that part of the total interaction term is
considered which commutes with the uncoupled evolution corresponding
to $H_0$.

\subsection{Decoupling with difference schemes}\label{harmonicadecouple}

The decoupling schemes presented in Section~\ref{network} can be used
to decouple general pair-interaction Hamiltonians. In this section we
consider a specific interaction which allows a more efficient
decoupling method using the combinatorial concept of difference
matrices (cf. \cite[Section VIII]{BJL1}).

The coupling $H_C$ between the harmonic oscillators can be removed by
transformations of the form $\exp( i h_k t)$ with $t\in \R$. If the
time evolution according to $H_C$ is conjugated by transformations
$\exp( i h_k t)$ and $\exp (i h_l s)$ on oscillator $k$ and $l$,
respectively, one part of the coupling term between $k$ and $l$ is
multiplied with the factor $\exp (i (s-t))$ and the adjoint part with
$\exp (i (t-s))$ since
\begin{equation}
\exp( i h_k t) \exp (i h_l s)\,  a_k a_l^\dagger\,  
(\exp(-i h_k t) \exp (-i h_l s)) =  
\exp (i (s-t))\,  a_k a_l^\dagger \, ,
\end{equation}
and a similar expression holds for the adjoint term.  

We now characterize $n\times N$ matrices $M$ having complex numbers of
modulus one as entries which are suitable for decoupling $H_C$. Such a
matrix $M$ defines a sequence of operations on the $n$ oscillators as
follows. If $(e^{i t_1},e^{i t_2},\dots,e^{i t_n})$ denotes the
$j^{\rm{th}}$ column, this means that during the $j^{\rm{th}}$ time
interval the natural time evolution of the $n$ oscillators is
conjugated by the local transformation
\begin{equation}
\prod_{k=1}^n \exp (i h_k t_k)\,.
\end{equation}
  
The total effect of this scheme is that the term $a_k a_l^\dagger$
obtains the factor $\langle m_l | m_k\rangle$, where $m_k$ is
the $k^{\rm{th}}$ row of $M$ and $\langle\cdot|\cdot\rangle$ is the
usual inner product in $\C^N$. This gives the following
\emph{decoupling criterion}: all couplings are removed if and only if
the rows are orthogonal in the usual sense. There is a canonical way
of finding $n$ vectors having this property by taking the Fourier
transform of the standard basis of $\C^n$. However, the rotations
$\exp(2\pi i h_j/n)$ required to be implemented are very close to the
identity for large $n$.

An alternative way of constructing such a matrix is given by
difference schemes \cite{BJL1,sloane}. We choose the numbers $t_k$ in
each row to be of the form $2 \pi r/u$ where $u\geq 2$ is a natural
number and $r$ is an element of the cyclic group
$Z_u=\{0,1,\dots,u-1\}$ of order $u$.

\begin{Definition}
An $n\times N$ array $D(n,N)$ with entries in $Z_u$ is called a 
difference scheme based on $Z_u$ if the difference vector of any two rows
has the property that each element of $Z_u$ occurs equally often.
\end{Definition}

Let $D(n,N)$ be a difference scheme based on $Z_u$.  We construct
an $n\times N$ complex matrix $M$ from $D$ by replacing each entry $r$
by $e^{2\pi i r/u}$. The rows of $M$ are vectors in $\C^N$ and since
$D$ is a difference scheme they are orthogonal, i.\,e.,
\begin{equation}\label{generalizedHadamard}
M M^\dagger = N\, \onemat_N\,.
\end{equation}
Therefore, $M$ satisfies the decoupling criterion. In view of
(\ref{generalizedHadamard}), a difference scheme $D(n,N)$ in which
$n=N$ is also called a generalized Hadamard matrix \cite[Section
VIII]{BJL1} of order $n$ over $Z_u$. In particularly, any (ordinary)
Hadamard matrix of order $n$ is a difference scheme $D(n,n)$ over
$Z_2$.

\subsection{Recoupling disjoint cliques}\label{Sel}
The scheme presented above does not only allow to remove all
interactions but also to achieve the following selective recoupling
without time overhead. Partition the set of nodes into $n'$ disjoint
subsets (called ``cliques'') and remove only the couplings between
nodes in different cliques.  This can be achieved by applying the same
sequences of transformations on all nodes in the same clique, since
this does not affect the interactions among them. Then it is
sufficient to construct a difference scheme with only $n'$ rows since
each row refers to one of the cliques.  Note that an analogous way of
``clique decoupling'' is also possible for the following kind of
$n$-qubit interaction. Assume that all qubits are coupled by the
interaction $\sigma_x\otimes \sigma_x +\sigma_y \otimes \sigma_y +
\sigma_z \otimes \sigma_z$. Then the interacting is invariant with
respect to simultaneous unitary rotations on both qubits. Hence
decoupling schemes for $n'$ qubits define a ``clique decoupling''
scheme for $n'$ cliques.

\subsection{Simulation of different coupling strengths}
An interesting problem arises if we want to switch off interactions
between arbitrary pairs or, even more general, to weaken the
interaction between some of the oscillators.  These goals are special
instances of the general problem to simulate a coupling Hamiltonian
$H_{\tilde{C}}$ using $H_C$ where $\tilde{C}$ is an arbitrary real
symmetric coupling matrix.

Now we define a matrix $T$ in such a way that the entry-wise product
(Schur-product) of $T$ and $C$ is $\tilde{C}$. Of course this is not
possible if an entry of $C$ is equal to zero and the corresponding
entry of $\tilde{C}$ does not vanish. This corresponds to the fact
that one cannot simulate a coupling between nodes which are not
coupled. We choose the vectors $m_k$ introduced in
Subsection~\ref{harmonicadecouple} with complex numbers of modulus one
such that the corresponding Gram matrix $G=(\langle m_k |
m_l\rangle)_{k,l=1,\ldots,n}$ coincides with $T$ on all off-diagonal
entries.  Note that the diagonal entries of $G$ always give the number
of time steps of the simulation. One can generalize this by choosing
time steps of different length and define the vectors $m_k$ as
mappings from the interval $[0,t]$ to the complex numbers of modulus
one. By taking the inner product
\[
\langle m_k | m_l \rangle := \int _0^t m_k(t') \overline{m}_l (t') dt'
\]
we obtain the same statement as above: the time overhead $t$ is
determined by the diagonal entries of $G$. Note that in our formulation a
weak interaction can be used to simulate a strong one since we allow
time overhead for the simulation.  

A \emph{lower} bound for this overhead is given by the absolute value
of the least eigenvalue of $T$ since $G$ being a Gram matrix is
positive. This coincides with the eigenvalue criterions stated in
\cite{graph} derived for $n$ spin $1/2$ systems. The lower bound for
simulating its own inverse has been shown to be $n-1$ for the
spin-spin interaction of the form $\sigma_z\otimes\sigma_z$
\cite{graph,arrow}. In strong analogy, we obtain the same bound $n-1$
for simulating $-H_C=H_{-C}$ by $H_C$ when $C$ has vanishing terms
only on the diagonal, since in this case the corresponding matrix $T$
has only entries $-1$ except for the diagonal entries which are zero.
This proves the bound since the absolute value of the least eigenvalue
of $T$ is $n-1$. Note that lower bounds based on the $T$-matrix
do only refer to simulations using unitary transformations
of the form $\exp(i h_j t)$ with $t\in \R$.
The following theorem shows that the lower bound $n-1$ for time reversal
is even valid if {\it general unitary transformations} on each oscillator
and presents an inversion scheme attaining this bound.

\begin{Theorem}[Optimal inversion]
A time optimal scheme for inverting a coupling
of the form $H_C$ as in equation (\ref{harmonicCoupling}) is given
by the vectors
\[
m_k:=(e^{2\pi ik/n}, e^{2\pi i2k/n}, \dots, e^{2\pi i (n-1)/n}).
\]
\end{Theorem}
{\bf Proof:} First note that due to the length of the vectors this
simulation has time overhead $n-1$. In order to prove that
this is optimal, define the following linearly independent elements of
$su(d)$.
\[
X_r:=|r\rangle \langle r-1| + |r-1\rangle \langle r|
\]
and
\[
Y_r:=i |r\rangle \langle r-1| -i|r-1\rangle \langle r|
\]
for $r=1,\dots, d-1$.  These matrices are orthonormal with respect to
the inner product $\langle V | W\rangle := {\rm tr}(VW)/2$.  One may
supplement these $2d-2$ vectors to a orthonormal basis of $su(d)$, but
the completion is irrelevant since the interaction among each pair of
oscillators can be written as an expression in $X_r$ and $Y_r$.
\[
a\otimes a^\dagger + a^\dagger \otimes a = \sum_{rs} \sqrt{rs}
(X_r\otimes X_s +
Y_r \otimes Y_s)\,.
\]
The coupling matrix $J$ (see eq. (\ref{Jmatrix})) can be constructed
as follows.  Define a $(2d-2)\times (2d-2)$-matrix $A$ by
\[
A:=|\phi \rangle \langle \phi | + |\psi \rangle \langle \psi|\,,
\]
with 
\[
|\phi\rangle := (\sqrt{1}, \sqrt{2},\dots, \sqrt{d-1}, 0,\dots,0)^T
\]
and 
\[
|\psi \rangle := (0,\dots, 0, \sqrt{1}, \sqrt{2},\dots, \sqrt{d-1})^T \,.
\]
With respect to the basis described above, for all pairs $(k,l)$ of
oscillators the coupling matrices $J_{k,l}$ are the same and given by
completing $A$ by embedding into a $(d^2-1)\times (d^2-1)$-matrix
$A'$.  Since the vectors $|\phi\rangle$ and $|\psi\rangle$ are
orthogonal, the spectrum of $J_{k,l}$ is $a,a,0,\dots,0$, where
$a=\langle \phi|\phi \rangle =\langle \psi |\psi \rangle >0$.  The
spectrum of $J$ contains two copies of the value $(n-1)a$ and $n-1$
times the value $-a$. The other eigenvalues are zero.  This can be
seen by writing $J$ as a tensor product $M\otimes A'$ where $M$ is an
$n\times n$-matrix with entries $0$ on the diagonal and $1$ elsewhere.
The smallest value $\tau>0$ such that the spectrum of $\tau J$
majorizes $-J$ is $n-1$. This proves optimality of the time reversal
scheme.  \qed

An \emph{upper} bound for the general simulation problem can also be
derived in strong analogy to \cite{graph} and is given by the
so-called {\it weighted chromatic index} $W_T$ of the matrix $T$. This
concept has been introduced in \cite{JB01} in a related context in
order to quantify the complexity of a general pair-interaction
Hamiltonian on $n$ qubits. For each $s\ge 0$ define a graph $G_s$ on
the $n$ nodes as vertices which has an edge $(k,l)$ if and only if the
absolute value of the entry $T_{kl}$ is greater than $s$. Let $n_s$ be
the chromatic number of $G_s$ (see \cite{bol}), i.\,e., the number of
colors required for coloring the edges of $G_s$ in such a way that no
two edges with a common node receive the same color. Then define $W_T$
as
\[
W_T:=\int_0^\infty n_s ds \,,
\] 
which defines a generalization of the chromatic index for weighted
graphs. The key idea to prove this upper bound can easily be
understood if one assumes $T$ to have entries of modulus one or
zero. Then $W_T$ is the chromatic index of a graph indicating which
couplings should not be removed. Given an admissible coloring of this
graph, we define a simulation with $W_T$ steps as follows. Each color
$c$ defines a step in which we remove all those interactions which are
not colored by $c$. This step can be executed without time overhead as
explained in Subsection~\ref{Sel}.  Hence in each step, only the
couplings between disjoint oscillator pairs remain.

%
%

\section{Comparison with other methods}\label{compare}

In this section we relate the decoupling method based on orthogonal
arrays which was presented in Section \ref{network} to the approach of
\cite{leung} for the case of decoupling in qubit networks (see also
\cite{stoll} for decoupling in qubit networks based on orthogonal
arrays).

Accordingly, let the system Hamiltonian be written in the form
\begin{equation}
  H = \sum_{kl;\alpha\beta} J_{kl;\alpha\beta} 
      \sigma_\alpha^k \sigma_\beta^l +
      \sum_{k;\alpha} r_{k;\alpha}\sigma_\alpha^k\,,
\end{equation}
where $\sigma_\alpha$ are the Pauli matrices.

We explain briefly the approach of \cite{leung} to construct
decoupling schemes. In each interval, each $\sigma^k_\alpha$ acquires
a $+$ or $-$ sign, which is controlled by the applied local unitaries
to be described. The coupling
$J_{kl;\alpha\beta}\sigma_\alpha^k\sigma_\beta^l$ for $k\neq l$ is
unchanged (negated) if the signs of $\sigma_\alpha^k$ and
$\sigma_\beta^l$ agree (disagree). Note that the signs of the three
Pauli matrices $\sigma_\alpha^k$ acting on the same qubit $k$ are not
independent.

Conjugating with the transformations
$\onemat^k,\sigma_x^k,\sigma_y^k,\sigma_x^l$ the acquired signs for
$\sigma_x^k,\sigma_y^k,\sigma_z^k$ are given by
$(+++),(+--),(-+-),(--+)$.

Following \cite{leung} a decoupling scheme for $n$ qubits that
concatenates $N$ intervals can be specified by three $n\times N$ sign
matrices $S_x,S_y,S_z$, related by the entry-wise product
$S_x*S_y=S_z$. We say that the three matrices satisfy the Schur
condition since this entry-wise product is usually called the Schur
product. The $(k,j)$ entry of $S_\alpha$ is the sign of
$\sigma_\alpha^k$ in the $j^{\rm{th}}$ time interval. Therefore
decoupling is achieved if any two rows taken from $S_x,S_y,S_z$ are
orthogonal.

The following theorem establishes a connection between decoupling
schemes constructed using orthogonal arrays as described in Section
\ref{network} and the decoupling schemes specified by sign matrices
$S_x,S_y,S_z$.

\begin{Theorem}
A decoupling scheme constructed using an orthogonal array $OA(n,N)$
over the alphabet $\mathcal{A}=\{1,2,3,4\}$ and the Pauli basis
$\cE=\{\onemat,\sigma_x,\sigma_y,\sigma_z\}$ for all nodes gives rise
to sign matrices $S_x,S_y,S_z$ satisfying the Schur and orthogonality
conditions.
\end{Theorem}
{\bf Proof:} We identify the operators of $\cE$ with the elements of
$\mathcal{A}$ according to $1 \mapsto \onemat$, $2 \mapsto \sigma_x$,
$3 \mapsto \sigma_y$, and $4 \mapsto \sigma_z$. Conjugating with
the operators of $\cE$ the Pauli matrices acquire the following signs:
\begin{equation}\label{sign}
\begin{array}{c|cccc}
         & 1 & 2 & 3 & 4 \\ \hline
\sigma_x & + & + & - & - \\
\sigma_y & + & - & + & - \\
\sigma_z & + & - & - & + \\
\end{array}
\end{equation}

Starting from the given orthogonal array we now construct the three
sign matrices $S_x$, $S_y$, and $S_z$. Pick any two rows $k$ and $l$
of the OA. We may assume that the two rows have the following form (or
else we apply a suitable permutation of the columns)
\begin{equation}
\bigg(
\underbrace{\begin{array}{cccc|cccc|cccc|cccc|c}
1 & 1 & 1 & 1 & 2 & 2 & 2 & 2 & 3 & 3 & 3 & 3 & 4 & 4 & 4 & 4 & \cdots\\
1 & 2 & 3 & 4 & 1 & 2 & 3 & 4 & 1 & 2 & 3 & 4 & 1 & 2 & 3 & 4 & \cdots
\end{array}}_{\lambda \mbox{ times }}
\bigg)
\end{equation}
since all pairs appear equally often ($\lambda$ times) in the OA.  Let
$\vec{\lambda}=(++\cdots+)$ be the vector of length $\lambda=N/16$
containing only $+$. By substituting the entries of the rows $k$ and the
$l$ of the OA by the corresponding sign assignments in
Table~(\ref{sign}) we define the following six rows of $S_x,S_y,S_z$,
respectively:
\begin{eqnarray*}
S_{x;k} & := & \vec{\lambda}\otimes (++--) \otimes (++++)\\
S_{y;k} & := & \vec{\lambda}\otimes (+-+-) \otimes (++++)\\
S_{z;k} & := & \vec{\lambda}\otimes (+--+) \otimes (++++)\\
& & \\
S_{x;l} & := & \vec{\lambda}\otimes (++++) \otimes (++--)\\
S_{y;l} & := & \vec{\lambda}\otimes (++++) \otimes (+-+-)\\
S_{z;l} & := & \vec{\lambda}\otimes (++++) \otimes (+--+)\,.
\end{eqnarray*}
Obviously, $S_x$, $S_y$, and $S_z$ are orthogonal and satisfy the
Schur condition $S_x * S_y = S_z$.  \qed

Finally, we give an alternative proof for the existence of a
decoupling scheme for $n=(2^{2m}-1)/3$ qubits using $N=2^{2m}$ time
intervals. A decoupling scheme with these parameters can be
constructed using orthogonal arrays \cite{stoll} and Hadamard matrices
\cite{leung}.

Let $V$ be the vector space $\mathbb{F}_4^m$, where $m \geq 1$ and let
$\mathbb{F}_4 = \{0,1,\omega,\omega^2=1+\omega\}$, where $\omega^3=1$,
be the Galois field with $4$ elements. Recall that the number of
$d$-dimensional subspaces of an $m$-dimensional vector space over
$\mathbb{F}_q$ is given by
\begin{equation}\label{subspaces}
\left[\begin{array}{c}m\\ d\end{array}\right]_q:=
\frac{(q^m-1)(q^{m-1}-1)\cdots (q^{m-d+1}-1)}{(q^d-1)(q^{d-1}-1)\cdots
(q-1)} 
\end{equation}
(cf.~\cite[Lemma 2.14, Section I]{BJL1}). For the special case $q=4$
and $d=1$ formula (\ref{subspaces}) shows that there are
$(4^m-1)/(4-1)=(2^{2m}-1)/3$ lines in $\mathbb{F}_4^m$. Note that
different lines intersect in the point $\{0\}$ only. Hence by taking
the set of all one-dimensional subspaces of $\mathbb{F}_4^m$ we obtain
a maximal spread in $\mathbb{F}_4^m$, i.\,e.\ a collection of
subspaces $U_i$ partitioning $\mathbb{F}_4^m$ with the additional
property that
\[
U_i\cap U_j = \{0\}\,. 
\]
We define a map $\varphi$ from $\mathbb{F}_4$ onto $\{-1,+1\}^4$ as follows:
\begin{eqnarray*}
\varphi(0)        & = & (+1,+1,+1,+1) \\
\varphi(\omega)   & = & (+1,-1,+1,-1) \\
\varphi(\omega^2) & = & (+1,+1,-1,-1) \\
\varphi(1)        & = & (+1,-1,-1,+1)
\end{eqnarray*}
This is the Hadamard matrix $H_4=H_2\otimes H_2$ where $H_2$ is the
usual Hadamard matrix of size $2$. Therefore all rows are
orthogonal.\footnote{The fact that this matrix is indeed the Hadamard
matrix can also be derived with the help of group characters. More
precisely, we consider $\mathbb{F}_4$ as a two-dimensional vector
space over $\mathbb{F}_2$ and let $\mathrm{tr}$ denote the trace map
of this field extension \cite[Section 4.15]{Jacobson:74}. For all
$z\in\mathbb{F}_4$ the map $\varphi(z) : x\mapsto
(-1)^{\mathrm{tr}(zx)}$ is an irreducible character of the additive
group $(\mathbb{F}_4,+)$ (which is isomorphic to $Z_2\times
Z_2$). Hence, orthogonality of the rows follows from the orthogonality
of the characters.}  Note that the last three rows satisfy the Schur
condition.

We extend the map $\varphi$ to vectors
$\vec{v}=(v_1,\ldots,v_m)\in\mathbb{F}_4^m$ by defining the map
\begin{equation}
\phi(\vec{v}):=\varphi(v_1)\otimes\cdots\otimes\varphi(v_m)
\in\{-1,1\}^{4m}\,.
\end{equation}
The image of $\phi$ is the set of all rows of the Hadamard matrix
$H_2^{\otimes 2m}$. Let $U_k=\langle\vec{v}_k\rangle$ be a maximal
spread of $\mathbb{F}_4^m$. By evaluating $\phi$ on the three elements
of $U_k$ (except for the zero vector $\vec{0}$) we get three
orthogonal vectors satisfying the Schur condition. We can take them as
rows of $S_x,S_y,S_z$:
\begin{eqnarray*}
  S_{x;k} & = & \phi(\omega\cdot\vec{v}_k)   \\
  S_{y;k} & = & \phi(\omega^2\cdot\vec{v}_k) \\
  S_{z;k} & = & \phi(1\cdot\vec{v}_k)
\end{eqnarray*}
This shows that the second through the last rows of the Hadamard
matrix $H_2^{\otimes 2m}$ can be divided into $(2^{2m}-1)/3$ disjoint
$3$-subsets, each with rows that satisfy the Schur condition.  The
rows in a $3$-subset can be chosen as rows of $S_x$,$S_y$ and $S_z$,
respectively.

\section{Conclusions}

We have shown that pair-interactions between the subsystems of a
multipartite quantum system can be decoupled efficiently if a
sufficiently large set of local control operations on the subsystems
is available. Such decoupling schemes can be constructed using {\it
orthogonal arrays} (a concept of combinatorics). The rows of
these arrays define pulse sequences of local operations taken from a
{\it unitary error basis}. We discuss the connection between the
decoupling method based on orthogonal arrays and those introduced in
\cite{leung}.

We have shown that mutual simulation of pair-interaction Hamiltonians
in multi node systems is possible provided that a so-called {\it
transformer group} of transformations is available.\footnote{Note that
similar results have been developed in \cite{Nielsen:2001},
independently.} The upper bound $O(n^2)$ on the simulation time is a
consequence of the existence of selective decoupling schemes. The
construction of a time-optimal simulation leads to a non-trivial
convex optimization problem. We have derived a lower bound on the time
overhead in terms of the spectrum of the matrix describing the
coupling parameters.  For some interactions simpler decoupling schemes
can be devised: for bilinear coupling of harmonic oscillators
selective decoupling can be achieved using so-called {\it difference
schemes}. The condition for time optimality of mutual simulation of
different bilinear couplings can concisely be expressed in terms of
linear algebra. Based on this we have constructed time optimal schemes
for time reversal.

From the results shown in this paper it follows that the time optimal
implementation of unitary transformations turns out not to be a matter
of optimal factorization into parallelized bilocal quantum gates
alone. The transformation has rather to be written as the solution of
a time-dependent Schr\"odinger equation where the occurring
Hamiltonians are those which can be simulated with small time
overhead.  This leads to another definition of {\it quantum
complexity} different from the discrete one measured by counting the
number of elementary gates.

\subsection*{Acknowledgments}

The authors acknowledge discussions with Markus Grassl. This work has
been supported by the European Community through grant IST-1999-10596
(Q-ACTA) and the DFG project {\em Komplexit{\"a}t und Energie}.

%
%

\bibliographystyle{plain} 
\bibliography{parallel}

\begin{thebibliography}{10}

\bibitem{two}
C.~H. Bennett, J.~I. Cirac, M.~S. Leifer, D.~W. Leung, N.~Linden, S.~Popescu,
  and G.~Vidal.
\newblock {Optimal simulation of two-qubit Hamiltonians using general local
  operations}.
\newblock Technical report, Los Alamos National Laboratory, 2001.
\newblock LANL preprint quant--ph/0107035.

\bibitem{BJL1}
{\kern0pt Th}.~Beth, D.~Jungnickel, and H.~Lenz.
\newblock {\em {Design Theory}}, volume~I of {\em Encyclopedia of Mathematics
  and Its Applications}.
\newblock Cambridge University Press, $2$nd edition, 1999.

\bibitem{bha}
R.~Bhatia.
\newblock {\em {Matrix Analysis}}, volume 169 of {\em Graduate texts in
  mathematics}.
\newblock Springer, 1996.

\bibitem{bol}
B.~Bollobas.
\newblock {\em {Modern Graph Theory}}.
\newblock Springer, Graduates Texts in Mathematics 184, 1998.

\bibitem{CD:96}
C.~J. Colbourn and J.~H. Dinitz.
\newblock {\em Handbook of Combinatorial Designs}.
\newblock CRC Press, 1996.

\bibitem{dodd}
J.~L. Dodd, M.~A. Nielsen, M.~J. Bremner, and R.~T. Thew.
\newblock {Universal quantum computation and simulation using any entangling
  Hamiltonian and local unitaries}.
\newblock Technical report, Los Alamos National Laboratory, 2000.
\newblock LANL preprint quant--ph/0106064.

\bibitem{ernst}
R.~R. Ernst, G.~Bodenhausen, and A.~Wokaun.
\newblock {\em {Principles of nuclear magnetic resonance in one and two
  dimension}}.
\newblock Clarendon Press, Oxford, 1987.

\bibitem{feynmann}
R.~P. Feynmann.
\newblock Simulating physics with computers.
\newblock {\em Int. J. Theor. Phys.}, 21:467, 1982.

\bibitem{sloane}
A.~S. Hedayat, N.~J.~A. Sloane, and J.~Stufken.
\newblock {\em {Orthogonal arrays: Theory and Applications}}.
\newblock Springer Series in Statistics, 1999.

\bibitem{Jacobson:74}
N.~Jacobson.
\newblock {\em Basic Algebra I}.
\newblock Freeman and Company, 1974.

\bibitem{JB01}
D.~Janzing and Th. Beth.
\newblock Complexity measure for continuous-time quantum algorithms.
\newblock {\em Phys. Rev. A}, 64(2):022301--1--6, 2001.

\bibitem{arrow}
D.~Janzing, P.~Wocjan, and Th. Beth.
\newblock {Complexity of inverting $n$-spin interactions: Arrow of time in
  quantum control}.
\newblock Technical report, Los Alamos National Laboratory, 2001.
\newblock LANL preprint quant--ph/0106085.

\bibitem{knill}
J.~A. Jones and E.~Knill.
\newblock {Efficient Refocussing of One Spin and Two Spin Interactions for NMR
  Quantum Computation}.
\newblock {\em J. Magn. Resonance}, 141:323--325, 1999.

\bibitem{KR:2000b}
A.~Klappenecker and M.~R{\"o}tteler.
\newblock Beyond stabilizer codes.
\newblock Technical report, Los Alamos National Laboratory, 2000.
\newblock LANL preprint quant--ph/0010076.

\bibitem{KR:2000a}
A.~Klappenecker and M.~R{\"o}tteler.
\newblock A remark on unitary error bases.
\newblock Technical report, Los Alamos National Laboratory, 2000.
\newblock LANL preprint quant--ph/0010082.

\bibitem{KCL:98}
E.~Knill, I.~Chuang, and R.~Laflamme.
\newblock Effective pure states for bulk quantum computation.
\newblock {\em Phys. Rev. A}, 57(3):3348--3363, 1998.

\bibitem{leung}
D.~W. Leung.
\newblock {Simulation and reversal of $n$-qubit Hamiltonians using Hadamard
  matrices}.
\newblock Technical report, Los Alamos National Laboratory, 2001.
\newblock LANL preprint quant--ph/0107041.

\bibitem{Lloyd96}
S.~Lloyd.
\newblock {Universal Quantum Simulators}.
\newblock {\em Science}, 273:1073--1078, 1996.

\bibitem{Nielsen:99}
M.~Nielsen.
\newblock Conditions for a class of entanglement transformations.
\newblock {\em Phys. Rev. Letters}, 83(2):436--439, 1999.

\bibitem{Nielsen:2001}
M.~Nielsen, M.~Bremner, J.~Dodd, A.~Childs, and C.~Dawson.
\newblock Universal simulation of hamiltonian dynamics for qudits.
\newblock Technical Report quant-ph/0109064, Los Alamos National Laboratory,
  2001.

\bibitem{vidalmaj}
M.~A. Nielsen and G.~Vidal.
\newblock {Majorization and the interconversion of bipartite states}.
\newblock {\em Quantum Information and Computation}, 1(1):76--93, 2001.

\bibitem{RPW:70}
W.-K. Rhim, A.~Pines, and J.~Waugh.
\newblock Violation of the spin-temperature hypotheses.
\newblock {\em Phys. Rev. Letters}, 25, 1970.

\bibitem{SOGKL01}
R.~Somma, G.~Ortiz, J.E. Gubernatis, E.~Knill, and R.~Laflamme.
\newblock {Simulating Physical Phenomena by Quantum Networks}.
\newblock Technical report, Los Alamos National Laboratory, 2001.
\newblock LANL preprint quant--ph/0108146.

\bibitem{stoll}
M.~Stollsteimer and G.~Mahler.
\newblock {Suppression of arbitrary internal couplings in a quantum register}.
\newblock Technical report, Los Alamos National Laboratory, 2001.
\newblock LANL preprint quant--ph/0107059.

\bibitem{vidal}
G.~Vidal and J.~I. Cirac.
\newblock {Optimal simulation of nonlocal Hamiltonians using local operations
  and classical communication}.
\newblock Technical report, Los Alamos National Laboratory, 2000.
\newblock LANL preprint quant--ph/0108076.

\bibitem{violaDecoupling}
L.~Viola, E.~Knill, and S.~Lloyd.
\newblock Dynamical decoupling of open quantum systems.
\newblock {\em Phys. Rev. Lett.}, 82:2417--2421, 1999.

\bibitem{graph}
P.~Wocjan, D.~Janzing, and Th. Beth.
\newblock {Simulating Arbitrary Pair-Interactions by a Given Hamiltonian:
  Graph-Theoretical Bounds on the Time Complexity}.
\newblock Technical report, Los Alamos National Laboratory, 2001.
\newblock LANL preprint quant--ph/0106077.

\bibitem{WRJB01}
P.~Wocjan, M.~R{\"o}tteler, D.~Janzing, and Th. Beth.
\newblock {Universal simulation of Hamiltonians Using a Finite Set of Control
  Operations}.
\newblock Technical report, Los Alamos National Laboratory, 2001.
\newblock LANL preprint quant--ph/0109063.

\bibitem{Zanardi00}
P.~Zanardi.
\newblock {Symmetrizing Evolutions}.
\newblock {\em Phys. Lett. A}, 258:77, 1999.

\end{thebibliography}

\end{document}